\documentclass[11pt]{asaproc}

\usepackage{graphicx}
\usepackage{subcaption}
\usepackage{amsfonts}
\usepackage{amsmath}
\usepackage{xparse}
\usepackage{courier}
\usepackage{natbib}
\usepackage{url}

\usepackage{times}

\title{ Large-Scale Online Experimentation with Quantile Metrics}

\author{Min Liu\thanks{LinkedIn Corp., 880 W Maude Ave, Sunnyvale, CA 94085, mliu@linkedin.com} \and  Xiaohui Sun\thanks{LinkedIn Corp., 700 E Middlefield Rd, Mountain View, CA 94043, xhsun@linkedin.com} \and  Maneesh Varshney\thanks{LinkedIn Corp., 700 E Middlefield Rd, Mountain View, CA 94043, mvarshney@linkedin.com} \and Ya Xu\thanks{LinkedIn Corp., 700 E Middlefield Rd, Mountain View, CA 94043, yaxu@linkedin.com} }
\begin{document}

\maketitle

\begin{abstract}
Online experimentation (or A/B testing) has been widely adopted in industry as the gold standard for measuring product impacts. Despite the wide adoption, few literatures discuss A/B testing with quantile metrics. Quantile metrics, such as 90th percentile page load time, are crucial to A/B testing as many key performance metrics including site speed and service latency are defined as quantiles. However, with LinkedIn's data size, quantile metric A/B testing is extremely challenging because there is no statistically valid and scalable variance estimator for the quantile of dependent samples: the bootstrap estimator is statistically valid, but takes days to compute; the standard asymptotic variance estimate is scalable but results in order-of-magnitude underestimation. In this paper, we present a statistically valid and scalable methodology for A/B testing with quantiles that is fully generalizable to other A/B testing platforms. It achieves over 500 times speed up compared to bootstrap and has only $2\%$ chance to differ from bootstrap estimates. Beyond methodology, we also share the implementation of a data pipeline using this methodology and insights on pipeline optimization.
\end{abstract}

\section{Introduction\label{intro}}
Online experimentation, also known as A/B testing\citep{box2005statistics,gerber2012field}, has grown in popularity across the technology industry as the gold standard for measuring impact. Many companies, Amazon, Facebook, Google, LinkedIn, Uber, to name a few\citep{tang2010overlapping, kohavi2013online, bakshy2014designing, xu2015}, have adopted this methodology and built in-house A/B testing platforms to streamline the A/B testing process and deliver experiment insights. 

At LinkedIn, A/B testing is at the core of data-driven decision making. Over the years, the A/B testing platform has evolved into an engine that powers testing needs across all produce lines, running hundreds of concurrent A/B tests daily, and reporting impacts on thousands of metrics per experiment\citep{Xu:2015:ICA:2783258.2788602, Xu:2018:SBS:3219819.3219875}. The fast product innovation cycle requires that the platform delivers reliable insights in a timely fashion, in specific, the first A/B testing report is generated less than 5 hours after experiment activation. Despite the large number of metrics reported in each experiment, all of them are average metrics, such as revenue per member or clicks per impression. This is because 1. average is a good enough summary statistic for most metrics, for example, optimizing for total revenue can be achieved through optimizing average revenue; 2. A/B testing with average metrics easily fits into the two sample t-test procedure\citep{deng2011choice}. There is one important type of metrics that cannot be nicely summarized by average, that is the performance metrics (e.g. page load time). Imagine two websites with exactly the same average page load time 0.5 second. Website A loads all pages in 0.5s while Website B loads $10\%$ pages in 5s and remaining $90\%$ pages in 0s. Despite the same 0.5s average page load time, Website A would be perceived as fast because each page loads within a blink of an eye, while Website B would be perceived as slow because users frequently need to wait for 5 seconds before a page loads. Therefore, to optimize the site speed experience for LinkedIn members, we need to reduce loading time of the slowest page loads, instead of reducing the average page load time by making the fast pages even faster. The industry standard for measuring page load time is \textbf{quantiles} such as \textbf{90-th percentile} or \textbf{p90}, and \textbf{50-th percentile} or \textbf{p50}. p90 monitors tail performance and is the ultimate performance metric to optimize for, while p50 monitors overall performance. Before implementing the quantile metrics A/B testing solution described in this work, average page load time was used as a surrogate for p50, but there was no good surrogate for p90, and experimenters did not have the capability to measure how their feature impacts members' site speed experience.


Enabling quantile metrics on the A/B testing platform unlocks many applications beyond measuring performance impact. It is useful whenever we are interested in the impact on the distribution of a metric, apart from a mere summary statistic of average. As one hypothetical example, an ecommerce website may be interested in growing total revenue without becoming overly dependent on a few popular items and losing bargaining power against suppliers of such items. They can achieve this goal by optimizing for average revenue per item, at the same time monitoring a few quantiles of revenue, such as p90, p50 and p20. As long as the quantiles are growing at a similar rate as the average, then the website is maintaining a good revenue balance among all items.

Despite of the importance of quantile metrics, no A/B testing platform is known to have enabled such testing capability prior to this work, primarily due the challenge with designing a solution that is both \textbf{statistically valid} and \textbf{scalable}. In order for A/B testing results to drive the correct decision, the impact estimate, statistical significance and error margin has to be statistically valid. Bootstrap\citep{efron1979} offers valid estimates, but is not scalable for the data size of LinkedIn  or most other tech companies; the asymptotic variance estimate assuming samples are i.i.d.\citep{rust1998} is scalable but ignores correlations among page load times, resulting in order-of-magnitude underestimation of p-value and exposing the experimenter to $61\%$ false positives when nominal false positive rate is $5\%$. In Section 2, we first describe both existing solutions and explain why they do not solve the quantile A/B testing problem, then we devote the remainder of Section 2 to presenting a statistically valid and scalable methodology for A/B testing with quantiles that is fully generalizable to other A/B testing platforms. It achieves over 500 times speed up compared to bootstrap and has only $2\%$ chance to differ from bootstrap estimates. In Section 3, we present numerical results comparing the proposed methodology to bootstrap in terms of statistical validity using 242 real experiments with different analysis population, date range, platforms (desktop, iOS and Android), page load mode, and quantiles (p50 and p90). There is only $2\%$ that the proposed standard deviation estimate differs from bootstrap, and when it does differ, the difference is below $7\%$, so when nominal false positive rate is $5\%$, the proposed methodology has an actual false positive rate at most $5.1\%$. Finally in Section 4, we outline the pipeline implementation and highlight the most important pipeline optimizations so readers who wish to build the same solution on their A/B testing platforms can easily apply similar optimizations.

\section{Methodology}
\subsection{Notations}
Suppose an A/B test is run with a number of variants\citep{Kohavi:2013:OCE:2487575.2488217}, where members in each variant gets a different experience. We are interested in measuring how the experience in each variant impacts the q-th quantile of page load time. In order to measure this impact and compute the statistical significance, we need estimates of the sample quantile and standard deviation of sample quantile in each variant. Zooming in on one variant, suppose in this variant there are:

Members $i = 1, 2, \dots, n$;

Member $i$'s page views $j = 1, 2, \dots, P_i$, where $P_i$'s are i.i.d. random variables following distribution $\mathcal{P}$;

Page load time of member $i$'s $j-th$ page view is $X_{i,j}$.

Suppose $X_{i,j} \sim \mathcal{F} $, but $X_{i,j}$'s are not necessarily independent of each other. In fact, page load times $X_{i,j}$ and $X_{i,j'}$ from the same member $i$ are likely positively correlated because page views from a member with fast device and fast network are likely to all be faster, and vice versa.

The $q-th$ sample quantile of $\{X_{i,j}, i = 1, 2, \dots, n; j = 1, 2, \dots, P_i\}$ is denoted $\hat{Q}$, and the variance and standard deviation of sample quantile are denoted $var(\hat{Q})$ and $stddev(\hat{Q})$.

\subsection{Existing Methodologies}
\subsubsection{Bootstrap}
\begin{figure}[h!]
    \centering
    \begin{subfigure}[t]{0.5\textwidth}
        \centering
        \includegraphics[width=0.9\textwidth]{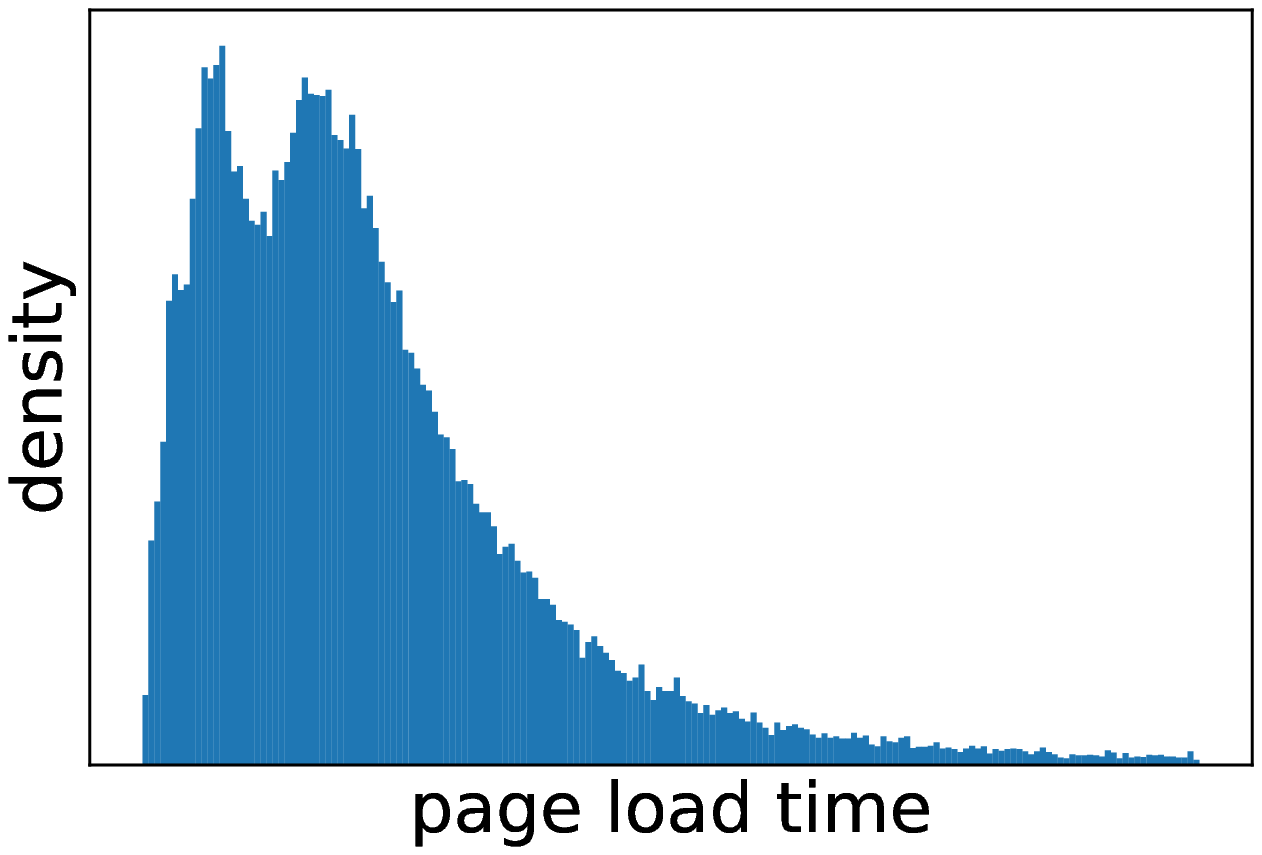}
        \caption{Distribution of non-i.i.d. page load times}
    \end{subfigure}%
    \hfill
    \begin{subfigure}[t]{0.5\textwidth}
        \centering
        \includegraphics[width=0.9\textwidth]{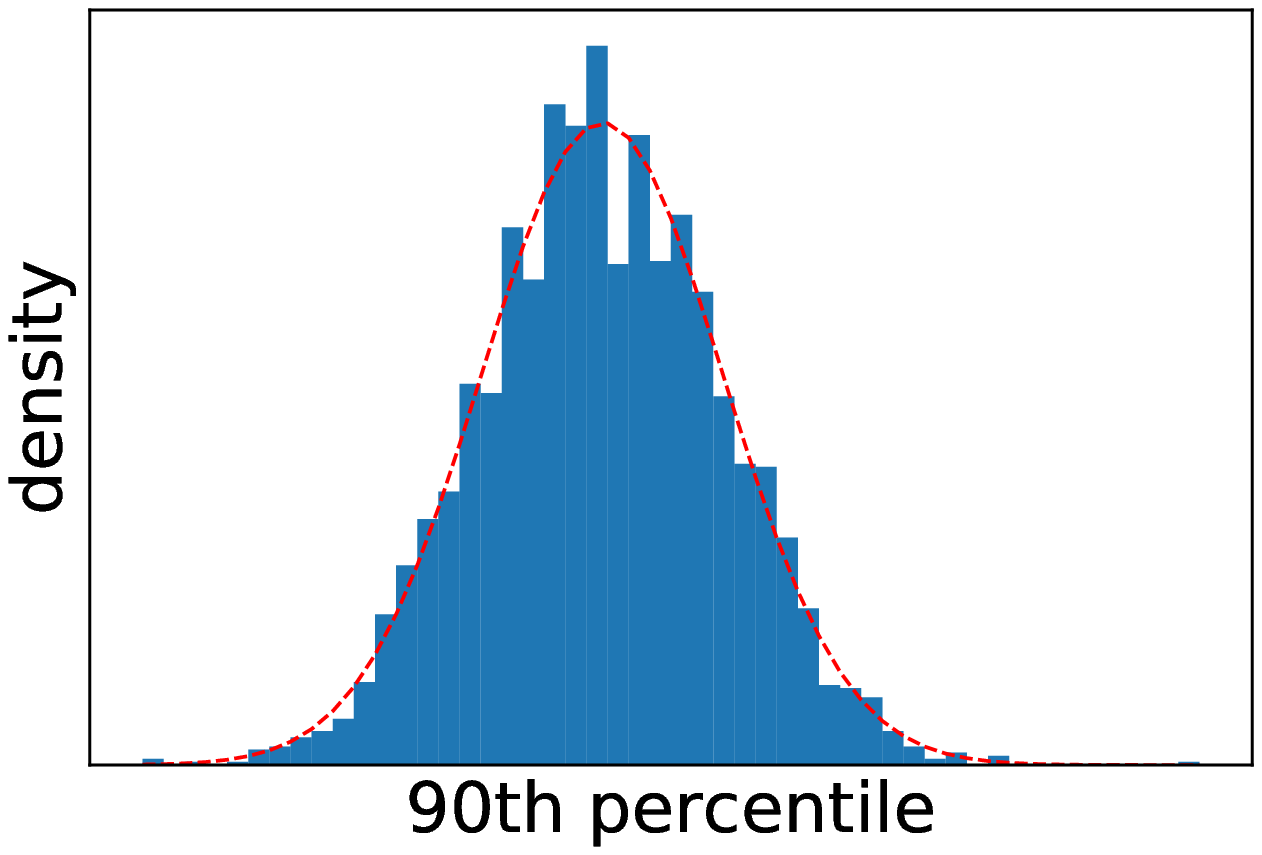}
        \caption{Distribution of bootstrap 90th percentiles}
        \label{bootstrap_density}
    \end{subfigure}
    \caption{Estimating standard deviation of sample quantile with bootstrap}
    \label{fig:bootstrap}
\end{figure}

Because page load times of the same member are not necessarily independent, but members are independent, the resampling in boostrap needs to happen on on member level to preserve the dependency structure. In the $k-th$ bootstrap sample, $n$ members are randomly sampled with replacement from the original $n$ members, then the $q-th$ sample quantile of the page load times of the $n$ resampled members are computed to be $\hat{Q}^{(k)}$. This process is repeated for $B$ times, and the sample mean and sample variance of $\{\hat{Q}^{(k)}; k = 1, 2, \dots, B\}$ are unbiased estimates of $\hat{Q}$ and $var(\hat{Q})$\citep{efron1979}. The sample standard deviation is a biased estimate of $stddev(\hat{Q})$, but the relative bias is on the order of $O(\frac{1}{n})$\citep{bolch1968}, so in a typical A/B test which has at least thousands samples, the bias is practically $0$. Figure \ref{fig:bootstrap} provides an example of the distribution of non-i.i.d. page load times and distribution of bootstrap 90-th percentiles, from which $stddev(\hat{Q})$ can be estimated. The red dotted line in Figure \ref{bootstrap_density} is the probability density function of a fitted normal distribution.

\subsubsection{Asymptotic Estimate Assuming Independence}
\begin{figure}[t]
    \centering
    \includegraphics[width=0.35\textwidth]{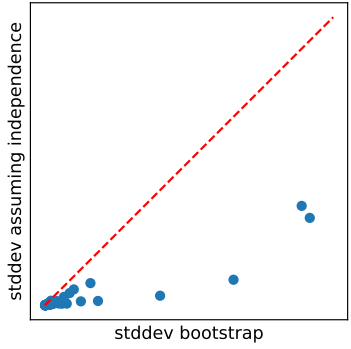}
    \caption{Underestimations of standard deviation with independence assumption}
    \label{fig:ind_vs_bootstrap}
\end{figure}

The asymptotic variance estimate for quantile of i.i.d samples is known \citep{rust1998}. If we apply this estimate on the page load time data assuming page load times are i.i.d. even though they are not, we can still get an standard deviation estimate. This estimate is, however, very much downward biased. See Figure \ref{fig:ind_vs_bootstrap} for how the asymptotic standard deviation estimate assuming i.i.d compares to bootstrap estimate, which is taken as ground truth given its unbiasedness. The median underestimation is $74\%$, which means when the estimated p-value is 0.05, the true p-value is actually 0.61, inflating the false positive rate by 12 times.

\subsection{Proposed Methodology}
Before delving into the details of the proposed methodology, it is worthwhile reiterating what is required of it: statistical validity and scalability. In order to make the correct data driven decision with A/B test results, the sample quantile and standard deviation estimates need to be valid; on the other hand, the fast product innovation cycle requires the pipeline be scalable enough to compute A/B test results from 300 billion rows of input data every day and finish computation in no longer than a few hours. A comparison of the methodologies is provided in Table \ref{methodology_comparison}.
\begin{table}[h!]
\caption{Comparison of Methodologies}
\begin{center}
\begin{tabular}{ccc}
\hline
\hline
\\[-5pt]
\multicolumn{1}{c}{Methodology} &
\multicolumn{1}{c}{Statistically Valid } &
\multicolumn{1}{c}{Scalable} \\
\hline
    Bootstrap & Yes & No \\
    Asymptotic estimate assuming independence & No & Yes \\
    Proposed Methodology & Yes & Yes\\
\hline
\end{tabular}
\end{center}
\label{methodology_comparison}
\end{table}

To establish a valid and scalable estimate for standard deviation of quantile of non-independent samples, we hope a closed form asymptotic distribution could be established through central limit theorem\citep{vandervaart}. The closed form expression would free us from the bootstrap and avoid the time consuming resampling process. The fact that the bootstrap quantile distribution in Figure \ref{bootstrap_density} matches well with a normal distribution strongly suggests such asymptotic distribution indeed exists.

The derivation is inspired by the asymptotic estimate assuming i.i.d.\citep{rust1998}, except here we do not make the unrealistic i.i.d. assumption, but only require that page load times from different members are independent, which is true whenever member is the randomization unit.

First we define $Y^{ \left( n \right) }_{} \left( x \right)  = \frac{1}{n} \sum _{i}^{n} \sum _{j}^{P_{i}} \mathbb{I}_{ \{ X_{i,j} \leq x \} }=\frac{1}{n} \sum _{i}^{n}J_{i}$ and $ P^{ \left( n \right) }_{ }=\frac{1}{n} \sum _{i=1}^{n}P_{i} $

 , where  $ J_{i}= \sum_{j=1}^{P_{i}} \mathbb{I}_{ \{ X_{i,j} \leq x \} } $. Naturally  $ J_{i}=0 $  if  $ P_{i}=0 $.

Under multidimensional central limit theorem,

\begin{equation}
    \sqrt{n}
    \left(
        \begin{pmatrix}
            Y^{ \left( n \right) } \left( x \right) \\
            P^{ \left( n \right) } \\
        \end{pmatrix}
        - 
        \begin{pmatrix}
            \mu _{J}\\
            \mu _{P}
        \end{pmatrix}
    \right)
    \xrightarrow{D}
    \mathcal{N} \left( 0,   \Sigma   \right)  
\end{equation}
 where $\Sigma$ is the variance-covariance matrix of $\left(Y^{ \left( n \right) }\left( x \right), P^{ \left( n \right) } \right)$, $\mu _{J}=\mathbb{E} \left(  \sum _{j=1}^{P_{i}} \mathbb{I}_{ \{ X_{i,j} \leq x \} } \right) =\mathbb{E} \left[ \mathbb{E} \left(  \sum _{j=1}^{P_{i}} \mathbb{I}I_{ \{ X_{i,j} \leq x \} } \vert P_{i} \right)  \right] = \mu _{P}F_{} \left( x \right)  $, $\mu_P = \mathbb{E}[P_i]$ and $ F \left( x \right)  $  is the cumulative distribution function of distribution $ \mathcal{F} $ .

Using the Delta method\citep{Oehlert1992},
\begin{equation}
    \sqrt{n} \left( \frac{Y^{ \left( n \right) } \left( x \right) }{P^{ \left( n \right) }}-\frac{ \mu _{J}}{ \mu _{P}} \right)   \xrightarrow{D} \mathcal{N} \left( 0,  \sigma _{P,J}^{2} \right)
\end{equation}

where  $  \sigma _{P,J}^{2}= \left( \frac{ \mu ^{J}}{ \mu ^{P}} \right) ^{2} \left( \frac{  \Sigma{JJ}}{ \left(  \mu ^J \right) ^{2}}+\frac{  \Sigma^{PP}}{ \left(  \mu ^P \right) ^{2}}-2\frac{  \Sigma  ^{PJ}}{ \mu ^{J} \mu ^{P}} \right) $ with $\Sigma^{JJ}$, $\Sigma^{PP}$ and $\Sigma^{PJ}$ elements in the $2\times2$ variance-covariance matrix $\Sigma$.

Let  $ F_{n} \left( x \right)  =\frac{Y^{ \left( n \right) } \left( x \right) }{P^{ \left( n \right) }}  $ , then the above expression can be written as,\par

\begin{equation}
    \sqrt{n} \left( F_{n} \left( x \right) -F \left( x \right)  \right)   \xrightarrow{D} \mathcal{N} \left( 0,  \sigma _{P,J}^{2} \right)
\end{equation}


When  $ x = \hat{Q} $ the $q-th$ sample quantile, that is, $q = F_n\left(\hat{Q}\right)$,

\begin{equation}
    \sqrt{n} \left( q -F \left( \hat{Q} \right)  \right)  \xrightarrow{D} \mathcal{N} \left( 0,  \sigma _{P,J}^{2} \right)
\end{equation}

Applying the Delta method again, because $ \hat{Q} $ is a consistent estimate of $ Q $ the population quantile,
\begin{equation}
    \sqrt{n} \left( F^{-1} \left( q \right)  - \hat{Q} \right) \xrightarrow{D} \mathcal{N} \left( 0, \frac{ \sigma _{P,J}^{2}}{f_{X} \left( Q \right) ^{2}} \right)
\end{equation}

Because  $ F^{-1} \left( q \right) =Q $  and the standardized normal distribution is symmetric, 
\begin{equation}
    \sqrt{n} \left( \hat{Q} - Q \right)  \xrightarrow{D} \mathcal{N} \left( 0, \frac{ \sigma _{P,J}^{2}}{f_{X} \left( Q \right) ^{2}} \right)
\end{equation}

\begin{figure}[h!]
    \centering
    \begin{subfigure}[t]{0.5\textwidth}
        \centering
        \includegraphics[width=0.9\textwidth]{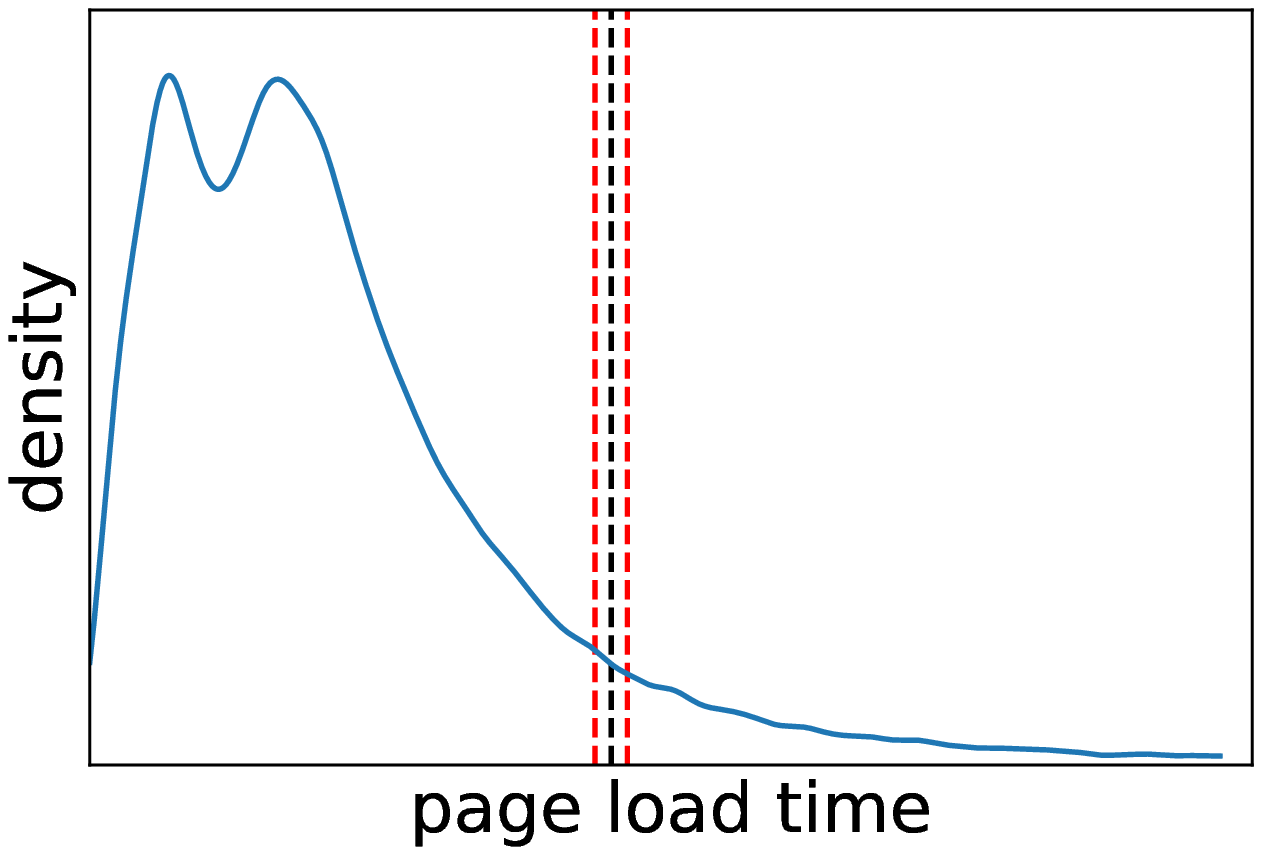}
        \caption{Probability density function of page load time}
    \end{subfigure}%
    \hfill
    \begin{subfigure}[t]{0.5\textwidth}
        \centering
        \includegraphics[width=0.9\textwidth]{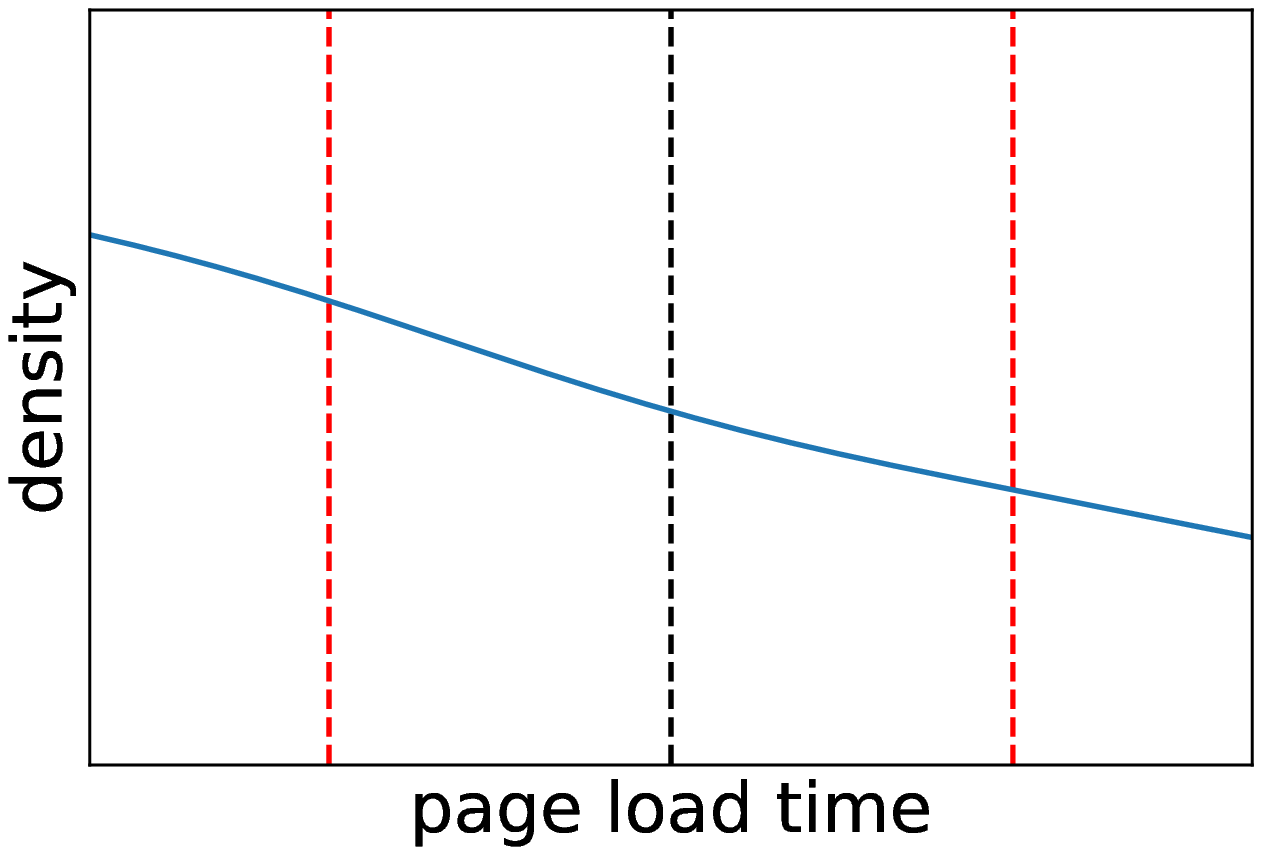}
        \caption{Local density around sample 90-th percentile}
    \end{subfigure}
    \caption{Density Estimation}
    \label{density_estimate}
\end{figure}

\begin{figure}[h!]
    \centering
    \includegraphics[width=0.35\textwidth]{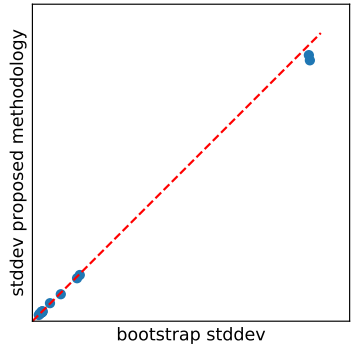}
    \caption{Accurate standard deviation estimate with proposed methodology}
    \label{proposed_vs_bootstrap}
\end{figure}

So the asymptotic estimate for variance of quantile is  $  \frac{ \sigma _{P,J}^{2}}{nf_{X} \left( Q \right) ^{2}} $, where the density at $Q$ can be estimated with the average density in a small interval aroud the sample quantile $\hat{Q}$(see Figure \ref{density_estimate}). The default interval size is set to $\pm 50ms$, which leads to a variance estimate that differs from bootstrap with roughly a one-in-ten chance. The estimate is worse with variance estimates for 90-th percentile than 50-th. This is expected as the density estimate is not bias free and could also be volatile especially far in the tail (e.g. at 90-th percentile) where there are not many data points around the sample quantile. The estimate can be very effectively improved by a dynamic interval width of $\pm 2\times stddev$, where $stddev$ is the standard deviation estimated in the first pass with $\pm 50ms$ interval. The dynamic interval width improves the estimate from $11\%$ error rate to only $2\%$. We have not proved mathematically why such dynamic interval width improves the estimate, but intuitively, a dynamic interval better balances bias and variance. When the standard deviation estimate is very large, $f(x)$ and $n_0$ are small, meaning there are few data points around the quantile, expanding the interval size from $50ms$ to $2\times stddev$ includes more data points and reduces the variance in density estimation. On the other hand, when the estimated standard deviation is very small, it means there are already a large number of samples in the interval, and we can reduce the interval size to reduce the bias in density estimation without increasing the variance much. An alternative approach we have tried is kernel density estimate, of which the interval estimate is a special case. Since the kernel estimate underperforms dynamic interval estimate and is also much harder to implement in the pipeline, we do not discuss it in this paper. A comparison between the proposed methodology VS. bootstrap is presented in Figure \ref{proposed_vs_bootstrap}, where the two estimates are almost identical, unlike the asymptotic estimate assuming independence in Figure \ref{fig:ind_vs_bootstrap}, which greatly underestimates the standard deviation.

One important observation that improved pipeline efficiency is that we only need members who actually have a page view to calculate the standard deviation. In triggered analysis\citep{kohavi2017online}, the experiment population includes any member who meets the trigger condition (e.g. visiting LinkedIn). However, not every one in this population has viewed the page (e.g. Jobs page) for which you intend to measure page load time impact. Here we show that in order to estimate the variance of quantile, you actually only need to process the members who had a page view on the page of interest, which greatly reduces storage and computation when the page has a low visitation rate.

Suppose out of $n$ members who triggered, only members $i = 1, 2, \dots, n_0$ had non-zero page views on the page of interest. Define $\mu^J_0 = \mathbb{E}\left(J_i\right | i=1,2,\dots, n_0)$, $\mu^P_0 = \mathbb{E}\left(P_i\right | i=1,2,\dots, n_0)$, $\Sigma_0 = Cov\left(J_i, P_i | i = 1,2,\dots,n_0\right)$, then

 $  \mu^{J}=\frac{n_{0}}{n} \mu^J_0 $ ,  $  \mu^{P}=\frac{n_{0}}{n} \mu^P_0 $

 $   \Sigma^{JJ}=\frac{n_{0}}{n}  \Sigma_{0}^{JJ}+\frac{n_{0}}{n^{}} \left( 1-\frac{n_{0}}{n} \right)  (\mu^J_0)^2 $ 

 $   \Sigma^{PP}=\frac{n_{0}}{n}  \Sigma_{0}^{PP}+\frac{n_{0}}{n^{}} \left( 1-\frac{n_{0}}{n} \right)  (\mu^P_0)^2 $

 $   \Sigma^{JP}=\frac{n_{0}}{n}  \Sigma_{0}^{JP}+\frac{n_{0}}{n^{}} \left( 1-\frac{n_{0}}{n} \right)  \mu^J_0 \mu^P_0 $

 $ \frac{1}{n} \sigma _{P,J}^{2}=\frac{1}{n} \left( \frac{ \mu ^{J}}{ \mu ^{P}} \right) ^{2} \left( \frac{  \Sigma^{JJ}}{ (\mu^J)^2}+\frac{  \Sigma^{PP}}{ (\mu ^P)^2}-2\frac{  \Sigma^{PJ}}{ \mu ^{J} \mu ^{P}} \right)$
 
 $=\frac{1}{n} \left( \frac{ \mu _{0}^{J}}{ \mu _{0}^{P}} \right) ^{2} \left( \frac{  \Sigma  _{0}^{JJ}}{\frac{n_{0}}{n} (\mu _{0}^J)^2}+\frac{1-\frac{n_{0}}{n}}{\frac{n_{0}}{n}}+\frac{  \Sigma  _{0}^{PP}}{\frac{n_{0}}{n} (\mu _0^P)^2}+\frac{1-\frac{n_{0}}{n}}{\frac{n_{0}}{n}}-2\frac{  \Sigma  _{0}^{PJ}}{\frac{n_{0}}{n} \mu _{0}^{J} \mu _{0}^{P}} - 2\frac{1-\frac{n_{0}}{n}}{\frac{n_{0}}{n}} \right)  $

 $ =\frac{1}{n_{0}} \left( \frac{ \mu _{0}^{J}}{ \mu _{0}^{P}} \right) ^{2} \left( \frac{  \Sigma  _{0}^{JJ}}{ (\mu _{0}^J)^2}+\frac{  \Sigma  _{0}^{PP}}{ (\mu _{0}^P)^2}-2\frac{  \Sigma  _{0}^{PJ}}{ \mu _{0}^{J} \mu _{0}^{P}} \right)  $

\subsection{Numerical Results}
In this section, we use 242 real A/B test datasets to evaluate standard deviation estimates using the proposed methodology VS. bootstrap. We can tolerate a $5\%$ difference in standard deviation estimate, since difference below this threshold cannot move a 0.04 p-value beyond 0.05, nor a 0.06 p-value below 0.05, therefore does not impact decision making. Any difference beyond $5\%$ is considered an estimation error. The A/B test datasets are chosen such that they contain a mix of different platform, geo-location, page load mode, page key, data range and quantile (see Table \ref{validation_variable}).

\begin{table}[h!]
\caption{Variables in Evaluation}
\begin{center}
\begin{tabular}{cccccc}
\hline
\hline
\\[-5pt]
\multicolumn{1}{c}{Platform} &
\multicolumn{1}{c}{Geo} &
\multicolumn{1}{c}{Date Range} &
\multicolumn{1}{c}{Page Key} &    
\multicolumn{1}{c}{Page Load Mode} &
\multicolumn{1}{c}{Quantile} \\

\hline
Desktop      &US&     1 Week  & Feed & Launch & 90th\\
iOS&     CN&      Weekend Only&        Jobs       & Subsequent & 50th\\
Android&     IN&   Weekend+Weekday   &  ... &       &\\
\hline
\end{tabular}
\end{center}
\label{validation_variable}
\end{table}

The evaluation results for desktop and mobile page load time quantiles are summarized in Table \ref{evaluation_desktop} and Table \ref{evaluation_mobile} at the end of the paper. Evaluation on estimates using both the fixed and dynamic interval widths are presented.

\begin{table}[b]
\caption{Evaluation Results Desktop}
\begin{center}
\begin{tabular}{ccccccc}
\hline
\hline
\\[-5pt]
\multicolumn{1}{c}{Page} &
\multicolumn{1}{c}{} &
\multicolumn{1}{c}{} &
\multicolumn{1}{c}{Date} &    
\multicolumn{1}{c}{Number of} &
\multicolumn{1}{c}{Errors} &
\multicolumn{1}{c}{Errors}\\
\multicolumn{1}{c}{Load Mode} &
\multicolumn{1}{c}{Geo} &
\multicolumn{1}{c}{Quantile} &
\multicolumn{1}{c}{Range} &    
\multicolumn{1}{c}{Experiments} &
\multicolumn{1}{c}{Fixed Interval} &
\multicolumn{1}{c}{Dynamic Interval}\\

\hline
INITIAL      & cn  & 50       & 1 week     & 2           & 0        & 0      \\
             &     &          & mix        & 2           & 0        & 0      \\
             &     &          & weekend    & 2           & 1        & 0      \\
             &     & 90       & 1 week     & 2           & 1        & 0      \\
             &     &          & mix        & 2           & 0        & 0      \\
             &     &          & weekend    & 2           & 2        & 0      \\
             & in  & 50       & 1 week     & 3           & 0        & 0      \\
             &     &          & mix        & 2           & 0        & 0      \\
             &     &          & weekend    & 2           & 0        & 0      \\
             &     & 90       & 1 week     & 3           & 1        & 0      \\
             &     &          & mix        & 2           & 1        & 1      \\
             &     &          & weekend    & 2           & 1        & 0      \\
             & us  & 50       & 1 week     & 3           & 0        & 0      \\
             &     &          & mix        & 4           & 1        & 0      \\
             &     &          & weekend    & 3           & 0        & 0      \\
             &     & 90       & 1 week     & 3           & 1        & 0      \\
             &     &          & mix        & 4           & 0        & 0      \\
             &     &          & weekend    & 3           & 1        & 1      \\
\hline
PARTIAL      & cn  & 50       & 1 week     & 4           & 2        & 0      \\
             &     &          & mix        & 4           & 0        & 0      \\
             &     &          & weekend    & 2           & 0        & 0      \\
             &     & 90       & 1 week     & 4           & 0        & 0      \\
             &     &          & mix        & 4           & 0        & 0      \\
             &     &          & weekend    & 2           & 0        & 0      \\
             & in  & 50       & 1 week     & 4           & 0        & 0      \\
             &     &          & mix        & 4           & 1        & 0      \\
             &     &          & weekend    & 4           & 1        & 0      \\
             &     & 90       & 1 week     & 4           & 0        & 0      \\
             &     &          & mix        & 4           & 0        & 0      \\
             &     &          & weekend    & 4           & 0        & 0      \\
             & us  & 50       & 1 week     & 4           & 0        & 0      \\
             &     &          & mix        & 4           & 0        & 0      \\
             &     &          & weekend    & 4           & 0        & 0      \\
             &     & 90       & 1 week     & 4           & 0        & 0      \\
             &     &          & mix        & 4           & 0        & 0      \\
             &     &          & weekend    & 4           & 0        & 0      \\
\hline
Total          &  &       &         & 114         & 14       & 2      \\
\hline
\end{tabular}
\end{center}
\label{evaluation_desktop}
\end{table}

\begin{table}[b]
\caption{Evaluation Results Mobile}
\begin{center}
\begin{tabular}{ccccccc}
\hline
\hline
\\[-5pt]
\multicolumn{1}{c}{} &
\multicolumn{1}{c}{} &
\multicolumn{1}{c}{} &
\multicolumn{1}{c}{Date} &    
\multicolumn{1}{c}{Number of} &
\multicolumn{1}{c}{Errors} &
\multicolumn{1}{c}{Errors}\\
\multicolumn{1}{c}{Platform} &
\multicolumn{1}{c}{Geo} &
\multicolumn{1}{c}{Quantile} &
\multicolumn{1}{c}{Range} &    
\multicolumn{1}{c}{Experiments} &
\multicolumn{1}{c}{Fixed Interval} &
\multicolumn{1}{c}{Dynamic Interval}\\

\hline
Android  & cn  & 50       & 1 week  & 3           & 2         & 0        \\
         &     &          & mix     & 3           & 1         & 1        \\
         &     &          & weekend & 3           & 0         & 0        \\
         &     & 90       & 1 week  & 3           & 0         & 0        \\
         &     &          & mix     & 3           & 1         & 1        \\
         &     &          & weekend & 3           & 0         & 1        \\
         & in  & 50       & 1 week  & 4           & 0         & 0        \\
         &     &          & mix     & 3           & 0         & 0        \\
         &     &          & weekend & 3           & 0         & 0        \\
         &     & 90       & 1 week  & 4           & 0         & 0        \\
         &     &          & mix     & 3           & 0         & 0        \\
         &     &          & weekend & 3           & 0         & 0        \\
         & us  & 50       & 1 week  & 4           & 0         & 0        \\
         &     &          & mix     & 4           & 0         & 0        \\
         &     &          & weekend & 3           & 1         & 0        \\
         &     & 90       & 1 week  & 4           & 1         & 0        \\
         &     &          & mix     & 4           & 0         & 0        \\
         &     &          & weekend & 3           & 1         & 0        \\
\hline
iOS      & cn  & 50       & 1 week  & 3           & 0         & 0        \\
         &     &          & mix     & 3           & 0         & 0        \\
         &     &          & weekend & 3           & 0         & 0        \\
         &     & 90       & 1 week  & 3           & 0         & 0        \\
         &     &          & mix     & 3           & 0         & 1        \\
         &     &          & weekend & 3           & 0         & 0        \\
         & in  & 50       & 1 week  & 4           & 1         & 0        \\
         &     &          & mix     & 4           & 2         & 0        \\
         &     &          & weekend & 4           & 0         & 0        \\
         &     & 90       & 1 week  & 4           & 0         & 0        \\
         &     &          & mix     & 4           & 0         & 0        \\
         &     &          & weekend & 4           & 1         & 0        \\
         & us  & 50       & 1 week  & 5           & 0         & 0        \\
         &     &          & mix     & 4           & 0         & 0        \\
         &     &          & weekend & 4           & 2         & 0        \\
         &     & 90       & 1 week  & 5           & 0         & 1        \\
         &     &          & mix     & 4           & 0         & 0        \\
         &     &          & weekend & 4           & 0         & 0        \\
\hline
Total      &  &       &      & 128         & 13        & 5        \\
\hline
\end{tabular}
\end{center}
\label{evaluation_mobile}
\end{table}

\section{Pipeline}
Now we shift gears to the engineering side. Figure 5 shows a high level flow of the pipeline. It is implemented in Spark and optimized to handle 300 billion rows of data. The main technologies used are: 1. data compression and data partitioning for parallel processing. 2. aggregate raw data into summary statistics within partitions to avoid data explosion\citep{Varshney2017}.

\begin{figure}[h!]
    \centering
    \includegraphics[width=0.9\textwidth]{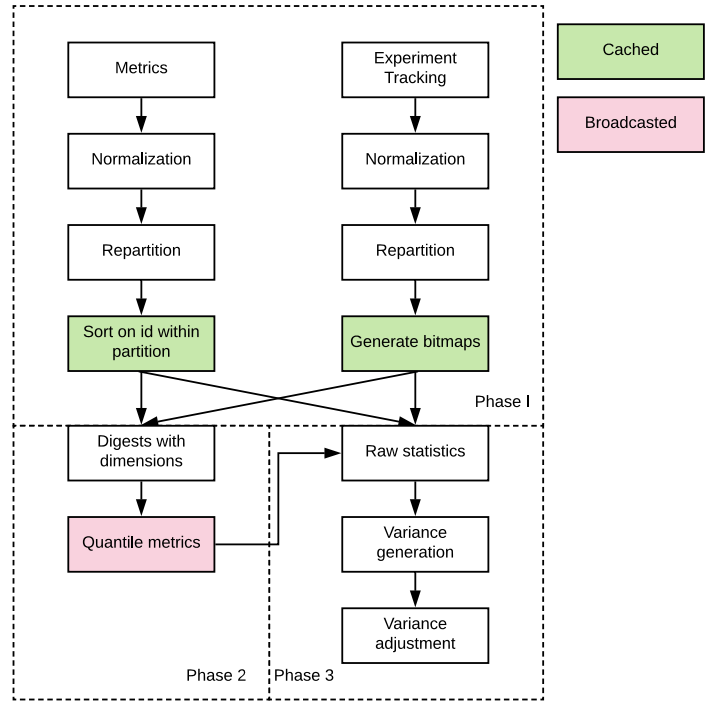}
    \caption{Quantile Computation Pipeline Workflow}
\end{figure}

The workflow takes two inputs:
\begin{enumerate}
    \item Metrics with schema \texttt{\{memberId, geo, platform/page load mode, page key, page load time, timestamp\}}. 
    \item Experiment tracking with schema \texttt{\{memberId, experimentId, segmentId, variant, timestamp\}}, that is which member participated in which experiment and variant on what day. 
\end{enumerate}{}
Outputs of the flow are quantile and variance of quantile for all combinations of \texttt{\{experimentId, segmentId, variant, geo, platform/page load mode\}}.

There are three phases in the calculation:
\begin{enumerate}
    \item Preprocess. Both metrics and experiment tracking are compressed and co-partitioned, the processed experiment tracking are further cached in memory to speed up subsequent joins.
    \item Quantile calculation. Metrics are joined with experiment tracking on memberId and timestamp using HashJoin, and quantile is calculated for all combinations of \texttt{\{experimentId, variant, geo, platform/page load mode, page key\}}.
    \item Variance calculation. This phase will take the quantiles computed in phase 2, and calculate variance for all combinations of \texttt{\{experimentId, variant, geo, platform/page load mode, page key\}}. 
\end{enumerate}

\subsection{Preprocess}
The preprocessing phase is composed of three steps:
\begin{enumerate}
    \item Normalization, which reduces the data storage size by encoding one or more columns into one integer index. For metrics, the geo, page load mode/platform and page key columns are combined and indexed; for experiment tracking, the experimentId, segmentId and variant columns are combined and indexed.
    \item Repartition. Co-partition the normalized metrics and experiment tracking by memberId and timestamp, so joining by memberId and timstamp can happen within partition, which reduces the complexity of join.
    \item Bitmap Generation. In this step the normalized experiment tracking data is transformed to a hash table of \texttt{(indexed \{experimentId, segmentId, variant\}, bitmap)}, where the bitmap holds memberIds of all members who were in \texttt{\{experimentId, segmentId, variant\}}. Bitmap further compresses the data and speeds up join by memberId and timestamp. The original experiment tracking data typically has over 4 billion rows every day therefore cannot be joint directly with metrics. On the other hand, the number of bitmaps is only on the order of thousands since there are only a few thousand combinations of \texttt{\{experimentId, segmentId, variant\}}. Therefore the bitmaps can easily fit in Spark memory and join with metrics efficiently.
\end{enumerate}

\subsection{Compute Quantile and Variance of Quantile}
The idea behind computing the quantile and variance of quantile are quite similar: first a summary statistic is computed within each partition, and then summary statistics across all partitions are merged to compute the quantile or variance of quantile. The only difference between the quantile and variance computation is that different summary statistics are computed. Producing summary statistics in each partitions reduces the amount of data merged across partitions and speeds up the flow.

The choice of summary statistic for quantile computation is essentially a histogram. In each partition, a histogram of page load times is produced for each combination of \texttt{\{experimentId, segmentId, variant, geo, platform/page load mode, page key\}}. Then histograms from all partitions are merged into the overall histogram from which any sample quantile can be computed. The summary statistics in quantile computation are $\sum_i J_i$, $\sum_i P_i$, $\sum_i J_i^2$, $\sum_i P_i^2$, $\sum_i J_i P_i$ and $\sum_i W_i$ where summation is over all members in the partition, and $W_i = \sum_j \mathbb{I}_{\{\hat{Q}+\delta \leq X_{i,j} \leq \hat{Q}+\delta\}}$ counts the number of page load times in an interval around the sample quantile, which is used to compute the density estimate.

The pipeline is able to compute 30 days of metrics and experiment tracking data, totaling in 300 billion rows, in an average of 2 hours.

\section{Summary and Future Work}
In this paper, we have presented a statistically valid and scalable methodology for A/B testing with quantile metrics, together with the pipeline implementation using this methodology. A detailed evaluation on real A/B test data shows the proposed methodology is over 500 times faster than bootstrap, and performs similarly in terms of statistical validity. Future work includes proving why dynamic interval width improved the variance estimation and research on more accurate density estimates.

\section{Acknowledgements}
We want to thank Nanyu Chen, Weitao Duan, Ritesh Maheshwari, Jiahui Qi and David He for insightful discussions and contributions to the implementation.

\bibliography{quantile_reference}
\bibliographystyle{apalike}


\end{document}